%% file: main.tex
\documentclass[sigconf]{acmart}

\AtBeginDocument{%
  }

\usepackage{xspace}
\usepackage{colortbl}

\acmYear{2026}
\setcopyright{cc}
\setcctype{by}
\acmConference[CHI '26]{Proceedings of the 2026 CHI Conference on Human Factors in Computing Systems}{April 13--17, 2026}{Barcelona, Spain}
\acmBooktitle{Proceedings of the 2026 CHI Conference on Human Factors in Computing Systems (CHI '26), April 13--17, 2026, Barcelona, Spain}
\acmDOI{10.1145/3772318.3791363}
\acmISBN{979-8-4007-2278-3/2026/04}

\input{macro}

\begin{document}


\title[How Generative AI Empowers Attackers and Defenders]{How Generative AI Empowers Attackers and Defenders Across the Trust \& Safety Landscape}

\author{Patrick Gage Kelley}\email{patrickgage@acm.org}
\affiliation{\institution{Google}\country{USA}}

\author{Steven Rousso-Schindler}\email{stevenrs@gmail.com}
\affiliation{\institution{CSU Long Beach}\country{USA}}
  
\author{Renee Shelby}\email{reneeshelby@google.com}
\affiliation{\institution{Google Research}\country{USA}}

\author{Kurt Thomas}\email{kurtthomas@google.com}
\affiliation{\institution{Google}\country{USA}}

\author{Allison Woodruff}\email{woodruff@acm.org}
\affiliation{\institution{Google}\country{USA}}

\renewcommand{\shortauthors}{Kelley et al.}

\begin{abstract}
Generative AI (\genAI) is a powerful technology poised to reshape \trust. While misuse by attackers is a growing concern, its defensive capacity remains underexplored. This paper examines these effects through a qualitative study with 43 \trust experts across five domains: child safety, election integrity, hate and harassment, scams, and violent extremism. Our findings characterize a landscape in which \genAI empowers both attackers and defenders. \genAI dramatically increases the scale and speed of attacks, lowering the barrier to entry for creating harmful content, including sophisticated propaganda and deepfakes. Conversely, defenders envision leveraging \genAI to detect and mitigate harmful content at scale, conduct investigations, deploy persuasive counternarratives, improve moderator wellbeing, and offer user support. This work provides a strategic framework for understanding {\genAIns}'s impact on \trust and charts a path for its responsible use in creating safer online environments. 
\end{abstract}




\maketitle

\input{01-Introduction}
\input{02-Background}
\input{03-Methodology}
\input{04-Findings}
\input{05-Discussion}
\input{06-Conclusions}

\begin{acks}
We thank our participants for generously sharing their insights and expertise. We thank Francesca Fenwick, Reena Jana, Tiffany Lin, Daniela Grafin Von Matuschka, Angela McKay, Xavier Morales, and Ashley Walker for their valuable contributions to this work.
\end{acks}

\bibliographystyle{ACM-Reference-Format}
\bibliography{ourbib}

\input{07-Appendix}

\end{document}

%% file: macro.tex
\newcommand\aff[1]{{\emph{--#1}}}
\newcommand\affpre[1]{{\emph{#1}}}

\newenvironment{listquote}{ 
  \setlength{\partopsep}{2pt}
  
  \begin{itemize}
    \fontsize{8.2pt}{8.2pt}\selectfont
    \setlength{\itemsep}{3pt}
    \setlength{\parskip}{3pt}
    \setlength{\rightskip}{15pt}
    \setlength{\parsep}{3pt}    
}{ \end{itemize} }

\newcommand{\genAI}{GenAI\xspace}
\newcommand{\genAIns}{GenAI} 
\newcommand{\GenAI}{GenAI\xspace}
\newcommand{\GenAIns}{GenAI} 

\newcommand{\trust}{Trust \& Safety\xspace}

 \def\Hype/{{\fontfamily{lmss}\selectfont\textbf{Sensationalized}}} \def\Academic/{{\fontfamily{lmss}\selectfont\textbf{Transformative}}}
 \def\Tool/{{\fontfamily{lmss}\selectfont\textbf{Effort-Saving Tool}}}
 \def\Narrative/{{\fontfamily{lmss}\selectfont\textbf{Narrative}}}
 
\newcommand{\eg}{e.g., }

\newcommand{\ie}{i.e., }
\newcommand{\etal}{et al.\@\xspace}

\definecolor{asparagus}{rgb}{0.53, 0.66, 0.42}
\definecolor{royalazure}{rgb}{0.0, 0.22, 0.66}
\definecolor{brickred}{rgb}{0.75, 0.25, 0.33}
\definecolor{bluegray}{rgb}{0.4, 0.6, 0.8}
\definecolor{darkmagenta}{rgb}{0.55, 0.0, 0.55}
\definecolor{darkslateblue}{rgb}{0.28, 0.24, 0.55}

\newcommand{\tagComp}[1]{    
  \newline\sffamily{\textbf{\textcolor{royalazure}{#1}}} }

\newcommand{\agendaG}[1]{\textcolor{bluegray}{#1}}



%% file: 01-Introduction.tex
\begin{center}
\setlength{\fboxsep}{.3cm}
\fbox{%
	\parbox{0.9\linewidth}{
		\textbf{Content Warning}: This paper discusses topics that some readers may find distressing, including child sexual abuse material (CSAM), hate speech, harassment, and violent extremism.
	}
}
\end{center}

\section{Introduction}
Recent advances in artificial intelligence have substantially raised expectations that it will transform work and society in areas such as economic productivity, scientific progress, education, and quality of life, among others~\cite{nationalacademies,ng2017,woodruff2024}. Many of these expectations stem from the newest \genAI systems, and their general-purpose nature, which allows them to be used for a wide variety of tasks across nearly every domain.

As with many areas, this technological shift is poised to broadly reshape the landscape of \trust, a field that has gained increased visibility in the CHI community in recent years~\cite{moran2025, cai2024}. \trust encompasses the organizations and disciplines focused on protecting the internet from myriad harmful activities and content, ranging from child safety to misinformation, through content moderation, platform policies, signal gathering, and investigations. The \trust field is characterized by an ongoing interplay between \textit{attackers} (\eg scammers, harassers) and \textit{defenders} (\eg content moderation teams, child safety hotlines, law enforcement). \GenAI may fundamentally change the threat landscape across \trust---who attackers are, who they target, how they operate, and the nature of their attacks. Concurrently, \genAI offers significant, underdeveloped opportunities for defenders to restructure their operations to counter both novel {\genAIns}-enabled attacks and longstanding ones~\cite{DTSP2024}. While prior research has focused on technical safeguards to minimize misuse of AI models and systems~\cite{shelby2023sociotechnical, weidinger2022taxonomy,amodei2016concrete, raji2023concreteproblemsaisafety,suresh2021,wei2023jailbroken}, these studies fall short of detailing how the uneasy sociotechnical balance between attackers and defenders will evolve alongside \GenAI.

We advance HCI's contribution to \trust by exploring how \genAI impacts the field, centering the perspectives of \trust experts on the front lines of fighting harmful content online. Drawing on participatory methods from HCI, we conducted six-hour participatory research workshops for five different \trust domains with a total of 43 expert participants from Asia, Europe, and North America. Our contributions include:
\clearpage
\begin{itemize}
\item We present a \textbf{novel qualitative study of experts in five \trust domains}\textemdash child safety, election integrity, hate and harassment, scams, and violent extremism\textemdash explor\-ing how \genAI currently affects their domains, and how they expect it to affect them in the future.
\item We describe a \textbf{dominant narrative that \genAI empowers both attackers and defenders}, a view shared in common by participants across all \trust domains studied.
\item We describe the ways in which participants expect \textbf{escalating competition between attackers and defenders} to unfold, with \genAI initially enabling attacks that outpace existing defense systems but subsequently enabling advanced defensive capabilities. At a high level, this view is shared in common across domains, but the specifics vary in different domains.
\item We identify \textbf{structural differences in different domains}, such as the nature of harm and the scale of attacker operations. Based on these differences, we draw practical implications for \trust operations by highlighting the need for domain-specific interventions.
\item Building on our participants' insights, we describe \textbf{the imperative, challenges, and opportunities for \trust organizations to transform their operations} to meet the novel threats posed by \genAI, and also, to fully realize the advantages of \genAI.
\end{itemize}

In the remainder of the paper, we review relevant background, describe our methodology, present our findings, and discuss tangible paths for creating safer online environments.

%% file: 02-Background.tex
\section{Background}
To situate our study of {\genAIns}'s impact on Trust \& Safety, we review the technical foundations of generative models, Responsible AI and Trust \& Safety, Trust \& Safety domains, and emerging literature on GenAI's effects in this area.

\subsection{\GenAI Models}
The most recent leap forward in artificial intelligence systems has been a set of technologies broadly referred to as \genAI, which includes large language models (LLMs) and other frontier models. These \genAI systems generate realistic or high-quality synthetic text, images, code, audio, and video, typically in response to a set of instructions called a prompt. Technologically, these systems are driven by an advancement in system architecture, the attention-based transformer model~\cite{vaswani2017attention}, which allows for training extremely large models on vast datasets using highly parallelized modern chip architectures. \GenAI systems are marked by three key characteristics: (1) applicability to generalized rather than specialized use cases; (2) production of original content that is often indistinguishable from human-created content; and (3) intuitive and accessible interfaces~\cite{briggs2023}. These systems are deployed as both proprietary models, typically equipped with safety guardrails by their developers, and as open-source models, which offer greater customizability and can be run with fewer inherent restrictions.

\subsection{Responsible AI and Trust \& Safety}
Responsible AI (RAI) is an umbrella term for research and practices aimed at developing safe, reliable, and ethical AI systems in a manner that aligns with societal values. Covering the entire system lifecycle, RAI is often structured around key principles~\cite{jobin2019global}, such as \textit{fairness}, to mitigate algorithmic bias; \textit{accountability}, to establish clear lines of responsibility; and \textit{transparency}, to ensure model decisions are understandable~\cite{floridi2022unified}. This upstream approach to technology development is enacted through governance structures and best practices for AI design, development, and deployment~\cite{shneiderman2020, rakova2021}. 

This applied work translates high-level RAI principles into concrete technical safeguards, focusing on preventing sociotechnical harms~\cite{shelby2023sociotechnical, weidinger2022taxonomy} by ensuring the robustness, reliability, and security of AI models and systems~\cite{amodei2016concrete, raji2023concreteproblemsaisafety}. It addresses vulnerabilities to prevent both direct misuse and harm from benign use~\cite{suresh2021}. This work involves a range of activities, including identifying social and ethical risks~\cite{rismani2023}; implementing pre-training interventions, such as improving the data the model learns from~\cite{suresh2021}; applying post-training mitigations to prevent undesirable behaviors like sycophancy or goal misinterpretation~\cite{perez2023}; and ensuring the fairness and robustness of tools used for content moderation, such as multimodal classifiers that detect hate speech~\cite{salminen2020} or other policy-violating content~\cite{gorwa2020algorithmic}. The field also focuses on mitigating risks such as ``jailbreaking,'' where users circumvent safety controls with adversarial prompts, and improving scaled evaluation techniques~\cite{wei2023jailbroken}.

Complementing these safeguards are the operational practices of Trust \& Safety, which strives ``to promote a safer and more trustworthy internet''~\cite{DTSP2025}. Historically, the field operated under a reactive model of incident response, such as removing user-flagged content or responding to legal requests after harm occurred~\cite{citron2025evolution, maxim2022build}; however, \trust is shifting toward a proactive strategy 
to embed safety into the design and architecture of a system~\cite{TPSA2021}. \trust roles vary significantly depending on a company's technology, user base, or the particular harms on which they focus. The Trust \& Safety Professional Association (TPSA) highlights diverse roles spanning policy design, threat analysis, and content moderation---many of which are similar to Responsible AI jobs~\cite{rismani2023RAI}. Professionals in these roles engage in a broad spectrum of activities, such as developing platform rules, supporting users, and moderating content~\cite{bhatlapenumarty2021}. The central aim is to mitigate risks and safeguard users from malicious actors, thereby enhancing user experience, reinforcing public trust, and meeting the complex legal requirements of various international frameworks.

\subsection{Trust \& Safety Domains}
Defense against online harms is not the sole responsibility of technology platforms but is instead managed by a diverse, multi-\allowbreak stake\-hold\-er ecosystem. This network includes platform Trust \& Safety teams, civil society organizations, government agencies, academic researchers studying emerging threats, and industry bodies that facilitate cross-company collaboration~\cite{DTSP2025}. The resilience of this ecosystem relies on effective coordination, as no single actor can address these complex challenges alone. 

Our work focuses on five key domains where this ecosystem confronts persistent, high-severity threats: child safety, election integrity, hate and harassment, scams, and violent extremism. We briefly describe these domains and some of the existing technologies used for detection and removal. A more exhaustive discussion of content moderation can be found in Singhal \etal~\cite{singhal2023sok} along with some of the professional challenges in Moran \etal~\cite{moran2025}. 

\emph{Child safety} is a foundational domain within Trust \& Safety, encompassing a wide spectrum of risks to minors, including grooming, cyberbullying, and exposure to age-inappropriate content. Among these harms, child sexual abuse material (CSAM) represents one of the most severe and legally unambiguous threats. CSAM involves the creation and distribution of real, photorealistic, or non-photorealistic sexual imagery of individuals under the age of 18. According to the International Center for Missing and Exploited Children (ICMEC), CSAM is illegal globally in all but 10 countries as of 2023~\cite{icmec2023model}, though definitions differ by jurisdiction. Historically, a relatively limited number of images---estimated at several hundred thousand---has circulated repeatedly~\cite{moderatedcontent}. Accordingly, technology platforms rely on perceptual hashes to match and report previously identified CSAM, with popular tools including PhotoDNA~\cite{photodna} and Google's Content Safety API~\cite{googlesafetyapi}. In the United States, the National Center for Missing and Exploited Children (NCMEC) serves as a clearinghouse for sharing CSAM hashes between platforms. NCMEC also receives reports of any detected content, with over 20 million such reports in 2024~\cite{ncmecvolume}. 
Longstanding challenges in this domain include the growing volume of content, the emotional toll on reviewers, and the risk of new material evading hashed-based detection~\cite{bursztein2019rethinking}. In 2023, NCMEC had its first ``million report day,'' due to a single viral meme which was manageable only with automated clustering, raising concerns that \genAI could enable a similar volume of unique images and overwhelm existing response systems~\cite{siotipline,thiel2023}.

\emph{Election integrity} covers threats such as false claims about voting procedures and fraud, intimidating or dissuading voters, and fabricated content related to candidates (\eg deepfakes). Concerns have surfaced regarding the use of \genAI by attackers, although some analyses suggest the harms to date have been more subtle than anticipated~\cite{motyl2024,myers2025,schiller2025}. Existing protections include fact checking partnerships, media literacy campaigns, and disclosure requirements for state-sponsored media.

\emph{Hate and harassment} refers to a broad spectrum of toxic content (\eg hate speech, trolling, sexual harassment) and maliciously leaked content (\eg doxxing, non-consensual explicit imagery) \cite{thomas2021sok}. Formal definitions of hate speech reflect just part of this spectrum, resulting in a fractured ecosystem of policies which is challenging for both moderators and users~\cite{riccio2024,shahid2023,stockinger2025}. Approximately 48\% of people globally report experiencing some form of online hate and harassment~\cite{thomas2021sok}. The scale, decentralization, and polymorphism of this content require solutions beyond hashing, most often platform-specific content classifiers (\eg the Perspective API~\cite{dixon2018measuring, wulczyn2017ex}) and user reporting.

\emph{Scams} are one of the broadest online threats, costing victims billions in losses annually~\cite{anderson2012measuring, cybercrime:weis15}. Examples include counterfeit products~\cite{levchenko2011click}, cryptocurrency giveaways~\cite{liu2024give, li2023double}, and ``pig butchering'' attacks~\cite{acharya2024explorative, oak2025}, among others. The profit incentive for scammers has led to an arms race in detection and evasion, with increasingly sophisticated scam infrastructure and content~\cite{zhang2021crawlphish}. Common detection strategies include content-based classifiers, account-based classifiers~\cite{xu2021deep}, and browser-based warnings~\cite{egelman2008you}, most of which are custom to each platform.

\emph{Violent extremism} (VE) glorifies, encourages, or facilitates violence to support ideological goals~\cite{baele2024, mitts2021}. Global definitions of VE and the specific entities it encompasses vary by country. Similar to CSAM, platforms rely on hash-based detection, with the Global Internet Forum to Counter Terrorism (GIFCT) acting as a hash-sharing clearinghouse. As of early 2024, GIFCT's database contained over 2.3 million hashes covering 408,000 distinct images, videos, and texts~\cite{gifctvolume}. Efforts to leverage \genAI for violent extremist and terrorist purposes have already been documented by several organizations~\cite{dsiegel2023,mathur2024,techagainstterrorism}.

\subsection{The Emerging Impact of \GenAI on Trust \& Safety}
A growing body of research and news articles has posited the risks \genAI poses in creating and distributing harmful content. Risks cited include promoting conspiracy theories~\cite{hill2025}, producing highly-believable fabricated media~\cite{myers2025}, increasing the quantity and quality of phishing scams~\cite{aiphishing}, generating synthetic, non-consensual sexually explicit imagery or CSAM~\cite{gibson2025analyzing, wei2025we, csamai}, or otherwise coercing users~\cite{matz2024, bartlett2025}. These risks are exacerbated by the possibility of attackers circumventing AI safety controls to produce harmful content~\cite{wei2023jailbroken, molas2024} and the inherent challenge of defining safety controls robustly~\cite{riccio2024, mchangama2024}. Our work expands on these perspectives by engaging experts across the Trust \& Safety ecosystem on the specific nuances of how \genAI will influence each of the five domains in our study.

Comparatively less attention has been paid by researchers and the media to the potential benefits of \genAI for improving defenses against harmful content. Prior studies have explored using AI to help debunk conspiracy theories~\cite{rosenbluth2024}, better detect scam content~\cite{bawa2025}, or scale the detection of hate and harassment, violent extremism, and election integrity threats~\cite{DTSP2024, stockinger2025, thomas2025}. Our work builds on these findings by exploring these potential applications from the perspectives of experts, identifying the opportunities and gaps they believe GenAI could help fill. This includes countering pre-existing threats, as well as new threats stemming from \genAI. While a significant body of literature exists for improving content moderation within specific \trust domains~\cite{singhal2023sok}, far less research has examined the cross-cutting challenges and opportunities that connect them. Furthermore, much of the existing technical work focuses on automated detection, with a dearth of research directly engaging the human experts who operate within these complex sociotechnical systems~\cite{moran2025}.

%% file: 03-Methodology.tex
\section{Methodology}
To examine the effects of \GenAI on \trust, we conducted participatory research workshops in Asia, Europe, and North America with experts from five domains: child safety, election integrity, hate and harassment, scams, and violent extremism. We draw on and extend the methodology presented in Woodruff et al.~\cite{woodruff2024}, which studied the impact of \genAI on knowledge workers. Our research examines the following questions:

\vspace{1cm}
\begin{itemize}
\item{How is \genAI already impacting \trust domains?}
\item{How is \genAI likely to impact \trust domains in the future?}
\item{How does {\genAIns}'s impact on \trust domains compare or interact with other changes in the field?}
\end{itemize}

\subsection{Participant Recruitment}
As \trust covers a wide range of digital safety harms, we selected five domains that have been highlighted in the \trust literature as top priorities~\cite{cryst2023}. In consultation with \trust experts, we confirmed these domains\textemdash child safety, election integrity, hate and harassment, scams, and violent extremism\textemdash vary in their harm profile, scope and scale, and regulatory environment, allowing us to elicit diverse perspectives across \trust issues. Furthermore, these domains are likely to be substantially affected by \GenAI~\cite{DTSP2024}. Considering multiple domains allowed us to interrogate existing operational and regulatory assumptions, which often take a one-size-fits-all approach to \trust harms management and mitigation. For more information about the domains, see Table~\ref{table:domain_overview}.

Following prior HCI and participatory design work examining how experts sensemake about novel technologies~\cite{sellen2002, millen2005, kawakami2023} and the impacts of technologies in their domains~\cite{nouwens2018, han2022, kawakami2024}, we engaged \trust workers as expert participants. Engaging experts is valuable because of their domain-specific knowledge, which offers essential insights regarding on-the-ground needs and norms~\cite{madaio2020, madaio2022, rismani2023, park2019} and situated perspectives towards potential sociotechnical harms~\cite{shelby2024}.

Accordingly, we recruited 8-10 workshop participants from each of the five domains, for 43 in total. For each workshop, we sought a mix of participants from civil society, industry, and academia. Within their domains, participants represented perspectives in areas such as education, human rights, journalism, law, law enforcement, machine learning, mediation, policy, and social science. Each participant received a personal invitation as a known expert in their field. We recruited participants in two ways: through mutual or personal connections or expressions of interest, and by identifying and contacting them based on relevant publicly visible expertise.

In this expertise-centered approach, participants and researchers collaboratively share and build upon each other's knowledge to reach a deeper understanding of a topic and envision solutions sensitive to the needs of multiple stakeholders (see Section~\ref{sec:workshopmethod}). In such an approach, it is neither expected nor necessary that experts fully grasp the underlying technical implementation of a particular technology to contribute meaningfully to the discussion of its impacts in their domains~\cite{hansen2020}, but rather, that they have expertise relevant to the research objectives. Thus, we prioritized recruiting participants with deep domain knowledge and on-the-ground experience rather than requiring specific technical knowledge of \genAI (although in fact many of the participants did have extensive experience with \genAI, especially regarding how it is used in their domains). Part of our role as researchers was to provide participants sufficient information about \genAI for them to relevantly engage, and it was also our role to gently address any potential misconceptions during discussion. Accordingly, we led a teaching intervention early in each workshop to bolster their understanding of \genAI (see Section~\ref{sec:workshopstructure}) and continued collaborative discussion of technological capabilities and characteristics throughout the sessions.

While our institution does not have an Institutional Review Board (IRB), we adhere to similarly strict standards. We received informed consent from all participants using a standard form, which included consent for a videographer to collect high-quality video and audio recordings. All participants were compensated at industry standard, receiving an equivalent amount in their local currency. A few participants who could not receive compensation directly due to institutional policies were offered the opportunity to direct the funds as a charitable donation. Additionally, travel reimbursements were available to most participants. For a summary of participants in each location, see Appendix A, Table~\ref{table:participants}.\footnote{In keeping with participant privacy practice discussed in Bellini et al., we provide limited detail about participants in order to minimize risk of identification~\cite{bellini}.}

\input{tables/domain_overview}

\subsection{Workshops as Research Method}
\label{sec:workshopmethod}
We conducted participatory workshops to engage with communities of practice representing five areas of \trust work. As a method, participatory workshops offer a structured way for researchers to gather ``reliable and valid data about [a] domain ... regarding forward-oriented processes, such as organisational change''~\cite[p. 73]{orngreen2017workshops}. This approach enables researchers to engage stakeholders to collaboratively explore their situated knowledge~\cite{harrington2019, angelini2025}, thereby grounding the inquiry in specific real-world contexts \cite{rosner2016}. In HCI, participatory workshops are a well-established methodology~\cite{ledantec2015, rosner2016}, particularly well-suited for engaging participants on emerging topics and capturing the ``real-timeness'' of a phenomenon, meaning actively engaging with the problem as it unfolds~\cite{darso2001}. They have been used to convene participants to reflect on domain-specific challenges~\cite{slovak2015, till2022}, discuss technologies relevant to their work~\cite{ahmadpour2025}, identify domain-specific harms~\cite{shelby2024}, and engage in speculative design~\cite{rosner2018, ratto2011}. We adopted this methodology specifically because of its alignment with critical theoretical traditions in HCI (\eg feminist, queer, and critical race perspectives) that recognize how technologies are differently experienced in ways shaped by situated norms and extant power dynamics~\cite{rode2011, bardzell2010, schlesinger2017, kumar2019}. This approach allowed us to: (1) center the direct, lived experiences of people affected by technology~\cite{rode2011, bardzell2010}; (2) examine technology as a driver of social change~\cite{light2011}\textemdash in our case, understanding the stages of \GenAI adoption in \trust work and exploring participatory views on potential sociotechnical and organizational interventions; and (3) employ research tactics designed to dismantle traditional hierarchies between researchers and participants~\cite{spiel2019, rankin2020, harrington2019}. To achieve these critical and contextual objectives, we designed our workshops to foster collaborative reflection, envisioning, and discussion, incorporating probes~\cite{boehner2007,gaver1999,bray2022,graham2008,beignon2020} and provocations to inform future agendas and policies around \GenAI in \trust.

\input{tables/agenda}

\subsection{Workshop Structure}
\label{sec:workshopstructure}
We held a total of five workshops, one per domain, between March 2024 and January 2025. We selected locations that are centers of activity for the respective domains, as well as being in different regions to represent a range of international perspectives. Furthermore, we held most workshops proximate to other large conferences relevant to the respective domains, in order to reduce travel burden for participants. Accordingly, we held workshops in Asia, Europe, and North America, with participants from a total of 15 countries (for specific cities and domains, see Appendix A, Table~\ref{table:participants}). Each six-hour workshop was held in person at a local Google office, and participants were aware of Google's involvement in the study. Each workshop was moderated by two or three researchers. An additional researcher, a visual ethnographer, attended each workshop in the role of videographer. The workshop agenda is shown in Table~\ref{table:agenda}.

In designing this six-hour agenda, we extended the three-hour workshop format detailed in Woodruff et al.'s study of knowledge workers~\cite{woodruff2024}. Our workshops retained two activities previously used successfully with knowledge workers: the \genAI teaching and Q/A, and the change cards activity~\cite{woodruff2024}. We customized these activities to the \trust context, leveraging our domain expertise, and we added three new activities using similar probes: the memorable experience cards activity, the attackers discussion, and the futuring stories activity. We piloted these new activities in the first workshop, and iteratively revised the futuring stories activity for the subsequent workshops as described below. The memorable experience cards activity and the attackers discussion required no modification.

We began by welcoming participants and setting expectations for the day. After a round of introductions, we facilitated a discussion about participants' memorable experiences with \genAI. For this purpose, we introduced a probe in the form of a small, physical card: a \textit{memorable experience card} (Appendix B,  Figure~\ref{fig:artifacts}). 
These cards encourage participants to reflect on any memorable experiences they have had with \genAI, whether related to their domain or not. Each participant spent about 10 minutes filling out their individual card(s). Participants then shared one or more of their memorable experience cards in a group discussion which we facilitated, allowing time for participants to react to each other's experiences. This activity introduced participants' orientation to and impressions of \genAI and laid the groundwork for future discussion. We then held a short break.

Next, a researcher from our team gave a presentation about \genAI, followed by a question and answer session, to provide participants with a shared understanding of this technology. The presentation focused on the key characteristics of \genAI as well as providing additional historical context (Appendix C). This presentation was followed by a meal break.

We then facilitated a discussion of potential upcoming changes to \trust. We began with a probe introduced in Woodruff et al.~\cite{woodruff2024}: a large, physical card called a \textit{change card}, which invites participants to share significant changes (related to \genAI, or not) that could take place in their domain in the next one to two years (Appendix B, Figure~\ref{fig:artifacts}). As described in Woodruff et al.~\cite{woodruff2024}, change cards support reflection and ``capture thoughts that participants [can] revisit and build on in collaborative discourse.'' Each participant spent about 10 minutes filling out their individual change card(s). Participants then shared and discussed the content of their change cards in an open-ended group discussion which we facilitated. Following this activity, we transitioned to an open-ended discussion of how participants believe attackers may use \genAI in their domains. We then held another break for refreshments.

Finally, inspired by futuring approaches~\cite{positech2024,positech2025,noh2025}, we invited participants to write an optimistic story set two years in the future about \genAI being used to effectively mitigate harm in their domains.\footnote{In the first workshop, we piloted a variant of this activity, seeded with post-it notes. Based on learnings from this activity, we settled on the final version described here for the four remaining workshops.} We encouraged participants to include classic storytelling elements such as setting, characters, plot, and/or conflict. We gave verbal instructions and provided each participant a large, physical \textit{futuring card} with the prompt, on which they wrote their stories (Appendix B, Figure~\ref{fig:artifacts}). Participants spent about 15 minutes working individually, and we included a checkpoint about halfway through this time to ask if participants had any questions or would like any support with their stories. After the individual work concluded, participants took turns reading their stories aloud in a facilitated discussion that lasted approximately 45 minutes, and we then concluded the session. Throughout the workshops we attended carefully to participant dynamics and supported collaborative discussion amongst all attendees.

\subsection{Data Preparation and Analysis}
The workshops produced three types of data: verbatim transcripts of the workshop discussions, participant-created artifacts (cards and stories), and researcher-produced synthesized video ``clip reels'' prepared by the visual ethnographer. Initial verbatim transcripts were created using an automated transcription service, and subsequently corrected for accuracy by the research team. This correction process involved reviewing the original recording while simultaneously reading and making necessary adjustments to the transcript. Identifying participant details were replaced with assigned pseudonyms (\eg, C1, C2) to maintain confidentiality. Participant-created artifacts, including 125 memorable experience cards, 99 change cards, and 28 futuring stories, were digitally scanned for analysis. For each workshop, the visual ethnographer created a video ``clip reel'' distilling the six-hour session into approximately 30 minutes of content highlighting participants' strong opinions and compelling stories aligned with the research questions. 

We undertook a reflexive thematic analysis~\cite{braun2019, braun2020} of this corpus, to inductively build themes centered on the impact of \genAI in \trust domains. The analysis involved deep and prolonged data immersion as well as discussion during meetings. Four members of the research team engaged in initial, independent open coding of the textual and video data, with concurrent analytical memoing to capture initial emergent themes~\cite{birks2008}. A designated researcher focused on open-coding the artifacts and integrating their findings into the shared analytical discussions. Throughout our analysis, the four researchers, including the visual ethnographer, held regular meetings to iteratively cross-compare codes and interpretations to reach consensus.\footnote{Consistent with practices commonly used for rich analysis of group interviews as described in McDonald et al.~\cite{mcdonald}, we focused on achieving researcher consensus through nuanced discussion rather than measuring inter-rater reliability.} The process continued until theoretical saturation of the data was reached, allowing us to finalize our main themes as well as create domain-specific thematic composites~\cite{creswell2018,willis2019}. The research team is composed of researchers with direct experience in both \genAI and the \trust domains studied, as well as related areas such as responsible AI development and AI governance. This experience allowed us to better interpret participants' comments, and assess alignment with consensus technical views of \genAI to validate the findings. Further, our experience allowed us to translate participants' insights about on-the-ground challenges and needs into actionable findings for \trust operations. Although the research team drew on our professional experiences, we engaged in a shared reflexive practice, continually discussing and interrogating the assumptions that arose during the analysis. 

\subsection{Limitations}
We invite readers to note several limitations of this study. Given that the research was conducted in only four international locations with single expert cohorts per domain, these findings offer a rich, contextually bounded perspective on situated expert knowledge but should not be taken as more broadly representative. In addition, the design of the workshops may have introduced a potential framing effect on participant commentary. This was counterbalanced, however, by prioritizing open, participant-led discussion and the participants' own pre-existing expert opinions. Finally, it is important to acknowledge that both the landscape of \genAI and expert understanding of it are changing rapidly. Therefore the expert consensus captured here reflects a snapshot in time, and views in \trust domains may shift as the technology matures.

%% file: tables/domain_overview.tex
\begin{table*}[h!]
\vspace{2.0em}
\centering
\begin{tabular}{ p{9em} p{4em} p{31em}  }

    Domain & Identifier & Brief Overview \\

    \toprule
    Child Safety & \emph{C} & \emph{Ensure child safety online. The discussion largely focused on preventing the creation and sharing of child sexual abuse material (CSAM) online.} \\

    \midrule
    Election Integrity & \emph{E} & \emph{Promote the integrity of elections, considering both misinformation and disinformation. The discussion encompassed civics integrity as well.} \\

    \midrule
    Hate \& Harassment & \emph{H} & \emph{Counter online hate \& harassment with approaches such as content moderation and content policy.} \\
    
    \midrule
    Scams & \emph{S} & \emph{Conduct anti-scam operations, to combat a wide range of online scams, such as romance scams or videoconference scams.} \\

    \midrule
    Violent Extremism & \emph{V} & \emph{Counter violent extremism and terrorist activities which are facilitated by online content and behavior.} \\

    \bottomrule

\end{tabular}
  \vspace{0.7em}
  \caption{Overview of each domain. In the Findings section, Identifiers are used to attribute quotations to domains.}
  \label{table:domain_overview}
\end{table*}

%% file: tables/agenda.tex
\begin{table*}[ht]
    \centering
    \begin{tabular}{l l l}
         \toprule
         \emph{Activity} & \emph{Time} \\
         \midrule
         \agendaG{Arrival, Breakfast} & \\
         \arrayrulecolor{bluegray}
         \midrule
         \arrayrulecolor{black}
         Welcome, Introductions              & \emph{30 minutes}     \\
         Memorable Experience Cards Activity & \emph{50 minutes}     \\  
         \agendaG{Mid-Morning Break}         & \agendaG{\emph{10 minutes}}    \\
         \GenAI Teaching and Q/A      & \emph{45 minutes}    \\
         \agendaG{Lunch Break}               & \agendaG{\emph{45 minutes}}    \\
         Change Cards Activity               & \emph{60 minutes}    \\
         Attackers Discussion              & \emph{30 minutes}    \\
         \agendaG{Mid-Afternoon Break}       & \agendaG{\emph{15 minutes}}    \\
         Futuring Stories Activity             & \emph{60 minutes}    \\
         Wrap-Up                             & \emph{15 minutes}    \\
         \bottomrule
    \end{tabular}
    \vspace{0.4em}
    \caption{Agenda for participatory research workshops. Times are approximate. }
    \label{table:agenda}
\end{table*}

%% file: 04-Findings.tex
\section{Findings}
\label{sec:findings}
Participants from all domains shared that \genAI can empower both attackers and defenders. Attackers can use \genAI for harmful purposes online. At the same time, defenders can use \genAI to implement safeguards against harm, whether it is perpetrated by \genAI or other means. For example, \genAI can enable both scam and anti-scam efforts.

\begin{listquote}
\item ``I think scams will continue to proliferate [within the next two to five years]... They get better, we get better. And the arms race keeps progressing.''~\aff{S6}\footnote{Throughout the paper, we assign participants pseudonyms that begin with the letter associated with their domain as shown in Table~\ref{table:domain_overview}, \ie \textit{C} (Child Safety), \textit{E} (Election Integrity), \textit{H} (Hate \& Harassment), \textit{S} (Scams), and \textit{V} (Violent Extremism). In some cases, we have lightly cleaned the quotes for readability, \eg to remove inconsequential false starts or to remove filler words such as ``like'' or ``um.''} 
\item ``Whilst clearly there's a huge threat in terms of the terrorist use of these technologies, there's also a significant opportunity in terms of using generative AI to understand, intervene, communicate, and so on and so forth [to counter terrorism]. So, incredibly excited by the benefits that generative AI will bring.''~\aff{V2} 
\item ``Bad actors will learn and will be able to use these tools to do whatever they want to do. But that's just what the tool is. It can be used for bad things as well as good things, and the good things are right next to it... If [bad actors] use generative AI to generate propaganda, that's something that can be countered by the opposite side. Like anti-extremists saying, `Okay, we can use the same tool [as extremists], but make good content.'{''}~\aff{C3} 
\end{listquote}

In this section, we describe participants' perspectives on the co-evolution of attackers and defenders in their use of \genAI. We begin by describing how \genAI can empower attackers, highlighting views held in common across domains. Next, we describe how \genAI can empower defenders, again highlighting views held in common across domains. We illustrate participants' concerns that \genAI will initially overwhelm existing, overtaxed response systems and that the defense ecosystem is reacting too slowly, as well as their optimism that \genAI can eventually be leveraged to counter both preexisting threats and new attacks facilitated by \genAI. Finally, we turn to unique perspectives within specific domains.

\input{04a-Attackers}
\input{04b-Defenders}
\input{04c-Domains}

%% file: 04a-Attackers.tex
\subsection{How Generative AI Empowers Attackers}
\label{sec:attackers}
Participants in all domains were concerned about ``all the innovative ways that bad actors are using AI'' 
and how \genAI can ``empower cyber criminals.'' 
Participants described memorable experiences with \genAIns-generated harmful content, highlighting current, active, damaging uses of \genAI. 

\begin{listquote}
\item ``I've seen obviously some sets of [CSAM] images or videos multiple times, as many of us probably have, and so I would say I know the faces of some of the victims of course... [I remember] the first time I think I saw an AI-generated image that was supposed to be photorealistic. But actually this one... they called it a ‘new set of images,’ of a victim that had been abused years ago, and she’s an adult now. Probably most of us know her actually. And I just saw these images and they looked almost photorealistic. This is when I thought, ‘Okay, this is an even worse step than just generating CSAM or other horrible content with AI or with other tools.’ And actually making new content of victims that had been abused [years ago] is a whole other level that was very different for me in that moment.''~\aff{C3} 
\end{listquote}

\noindent They anticipated even more harm in the future, given the sophistication and rapid rate of adoption of \genAI tools. They emphasized that existing \trust operations are not equipped to meet these ``alarming'' and ``terrifying'' new threats.
In the remainder of this section, we describe how scale and speed, lowered barriers to entry, enhanced messaging, degradation of the information environment, and harmful personal relationships with chatbots or AI companions empower attackers.

\subsubsection{Scale and Speed}
Participants explained that attackers are using \genAI to scale up harmful online content and behavior in ways that outpace existing defense mechanisms. Participants highlighted that \genAI dramatically increases the volume and speeds up the production of harmful content in every \trust domain studied, and as a result, ``all [harmful] content is just exploding in volume.'' 

\begin{listquote}
\item ``AI CSAM has dramatically risen as a new category...''~\aff{C9 (Memorable Experience Card)} 
\item ``... not being able to keep up with the scale of content moderation, because now it's just so much easier, faster, cheaper to do any of it. It's so much cheaper than renting a troll farm or something. And you can make it a lot more targeted and personalized... Content moderation was already really, really hard and something that didn't always get enough time or resources. And it just got so much harder.''~\aff{H2} 
\end{listquote}

Taking scams as an example, using \genAI can be easier and faster than perpetrating them manually. For example, if a romance scam involves sending text messages over a prolonged period, \genAI can create and send those messages. Consequently, where a single individual might previously have spent time messaging a small number of people, they can now oversee dashboards of \genAI chatbots sending messages to a much larger number. This pattern, where a single actor can multiply their impact, was a common concern across all domains.

Accordingly, participants emphasized the strain \genAI places on existing content review, legal, and investigative systems. They saw \genAI attacks as taxing an already under-resourced and under-prioritized \trust ecosystem, with some describing the status quo as a failed system. In child safety, participants observed that existing mechanisms such as hashlists would need to evolve both technically and operationally to defend against new challenges posed by \genAI. Furthermore, participants expressed concern that the defender ecosystem is adapting too slowly to challenges raised by rapidly evolving \genAI technologies. At the same time, they worried that distant future threats receive disproportionate attention at the expense of urgent, ongoing problems.

\begin{listquote}
\item ``We’re already seeing increasing workload, all of us. Hugely increasing. Not like ten percent a year, more like sometimes a couple of hundred percent more a year, in reports that we get, in content that we have to assess and work on. On the other hand, our resources are not exploding like that. They’re more stagnating.''~\aff{C3} 
\item ``What we have seen pretty consistently [so far] is that it’s the use of broadly shared open-source models that are manifesting in these kinds of environments. They're still mostly diffusion models... When I think about detection, one of the things that I'm concerned about is [attackers will] eventually find some newer, more exciting architecture than diffusion models to produce these [CSAM] materials, and then all of the detection technologies we’ve built will now be out of date and will need to be rebuilt in order to address that problem.''~\aff{C10} 
\item ``Oftentimes our discussions of what AI can do end up focusing too much on the speculative... I think I focus too much on, ‘Hey, what's gonna happen in the next few years?’ when the harm that is happening right now is very apparent...''~\aff{V4} 
\item ``These are actually really hard problems and what's really concerning is that there's a lot of time and effort and resources and money just being spent in trying to figure out what we're gonna build in the next 10 years, and how we're going to get AGI or what that will look like, versus, `Let's solve the problems that we have today.' There's a lot of them, and there just isn't enough time or focus or just money being spent on that.''~\aff{H2} 
\end{listquote}

\subsubsection{Lowered Barriers to Entry}
Participants expressed significant concern that the widespread availability and ease of use of \genAI tools have lowered the barrier to entry for creating and propagating harm. Readily available \genAI tools require relatively little technical skill, making it easier to create harmful content and widening the range of potential bad actors. For example, participants raised concerns about the rise of AI nudification apps and their use by teens. Similarly, just as specialized software toolkits previously enabled more people to engage in scams, participants anticipate that \genAI tools will pave the way for more attackers to enter the space. Participants noted from experience that mainstream models are often easy to ``jailbreak'' with little technical skill, and open-source tools with reduced safeguards further exacerbate the situation.

\begin{listquote}
\item ``The barrier to entry's been lowered. So it's easier, faster to [scam people], and you don't need that much skill. They're doing it more because now it's easier for them to do it, and it costs less.''~\aff{S9} 
\item ``If we have GenAI, when the barrier is so low, we can see more individual fraudsters coming out. Even the people who were not fraudsters may say, ‘But, oh, it's so easy... Why don’t I just do it?’{''}~\aff{S1} 
\item ``[In my experiments with jailbreaking] if I want guidance on how to craft propaganda, what tools I'd like to use to influence a given population, I can get those answers. If I want bomb making, I can get those answers. If I want CBRN [Chemical, Biological, Radiological, and Nuclear] ideas, I can get those. It's really remarkable.''~\aff{V1} 
\item ``They're playing with it... Right now they're in the stage where they're still experimenting... and they're doing the same things that we're doing, which is, [what can I do] on a chatbot? And now [that more safeguards are in place in commercial chatbots], they're escalating to open source models.''~\aff{H5} 
\end{listquote}

Some adversarial organizations are well-organized and well-resourced. For example, nation-state actors may be involved in violent extremism or election mis/disinformation, and organized crime organizations often operate scams. \GenAI can empower these types of sophisticated attackers. At the same time, attackers also include individual or loosely organized actors, such as online CSAM communities sharing tips for using models to produce harmful content. Participants believe \genAI can significantly elevate the capabilities of these less well-resourced attackers.

\begin{listquote}
\item ``[We’ve] started seeing how collaborative bad actors are being... that they're working together to fine tune these models with existing CSAM material to produce more bespoke content... They're working together to try to figure out how to best misuse these models, how to fine tune them, giving each other technical tips on what that looks like.''~\aff{C10} 
\item \affpre{H5}:\footnote{When presenting dialogic sequences that represent conversational exchanges between participants, we use script format prepending each utterance with the participant's identifier.} We are talking about these larger groups, like the well-funded ones, but that knowledge eventually trickles down... abusive actors get together and share this knowledge like, ‘Oh, did you see?’ They're talking about it, like, what's happening on Twitter, and they're trying to replicate it. They don't have the knowledge yet. They don't have the money yet, but that knowledge is gonna trickle down eventually. \newline 
\affpre{H2}: I think it makes it so much easier for even lone wolfs to do it now, right? You don't actually need a really big operation to do something like this now. 
\end{listquote}

In addition to widening the range of potential attackers, increased accessibility and ease of use also widens the range of potential targets, often to people with fewer resources to defend themselves. Participants explained that \genAI empowers both sophisticated groups and those with fewer resources to inflict harm on a wider range of targets.

\begin{listquote}
\item ``We're entering a space in which anybody can be victimized. Like, you have one photo, one video of yourself online, and you can be victimized in this way. And there will be different levels of innate protections in place depending on your current posture and status in society. Maybe a public figure can have some assumptions that they would have a faster response, that [a platform] is going to move on it because it's Taylor Swift. Will that similarly happen if it's just somebody else? Unlikely... That’s really going to leave a lot of people in a terrible, unsupported spot.''~\aff{C10} 
\item \affpre{H2}: Now you can see the general public getting targeted... [more people who are] not celebrities or known figures. Now you can see it with high school teachers, or high school students, or children, because now you can do that. You can get an image of them and you can change it to whatever you want to. So I think there has been kind of a shift there. \newline 
\affpre{H7}: Yeah. I feel like before, you had to use Photoshop... you had to have expertise, you had to have time, and all of that type of stuff. And now you don't need any of that. The barrier of entry is so low. 
\end{listquote}

\subsubsection{Enhanced Messaging}
Participants described how attackers are using \genAI to create and spread sophisticated harassment campaigns or propaganda, such as disinformation and extremist narratives. Not only can \genAI produce polished, professional content, from individual messages to networks of websites, it can tailor it to specific audiences or individuals. Further, \genAI enables attackers to generate content in languages they do not speak, allowing them to target new populations. In these ways, \genAI content can reach a wide audience and confer an appearance of legitimacy, adding more depth and background to attacks that might otherwise have been shallow.

\begin{listquote}
\item ``The propaganda use cases are very interesting. It's not that it changes the entire paradigm, but it changes the speed and the ease of access to readily aligned types of content that will meet the needs of your audience.''~\aff{V1} 
\item \affpre{S9}: [Before you would] translate from English to Chinese, for example, in one language at a time. But with \genAI, you can ask it to generate disinformation in ten different languages simultaneously. You can ask it to generate deepfake videos of some president saying something that he didn't really say, in nine different languages at the same time. And it's with one prompt. So it’s much easier. \newline 
\affpre{S3}: Yeah. It's the ability to chain link the different things, right? So you can generate in ten languages, create websites of those, make phone calls, and then get people to come to the websites. I can do all those things together on one platform. \newline 
\affpre{S4}: Yeah, it's really the scale to do it across different platforms as well, right? Like I can send SMS to a thousand people now easily, customized for different languages or context. And the links would be up online for the websites. 
\end{listquote}

Participants emphasized \genAI-enabled creation of highly entertaining terrorist and violent extremist content designed for widespread appeal to audiences on social media and other large platforms, \eg memes, anime, or funny images of violent attacks or harmful ideological content. This phenomenon makes it difficult to distinguish an actual extremist from someone who is simply reposting engaging content (perhaps without realizing its import), which V6 noted makes it difficult to ``put policy lines around.'' Similarly, V4 explained that entertaining content can ``evade moderators.'' More broadly, participants expressed concern that malign actors are learning to share harmful materials that are not policy-violative and/or do not get flagged by content moderation tools. This can include entertaining content such as that described above, coded messages, or other apparently benign content. \GenAI can greatly simplify the production of such material, which is often designed to outlink from large platforms to funnel users to more specific terrorist content, tools, or communities, providing an escalation path for recruitment.

\begin{listquote}
\item ``Something that I've seen that's pretty tough to deal with, particularly from a platform perspective, and it's relatively easy to do too, is using voice clones to take prominent individuals in social media or in television or in movies, to take their voices and then game the algorithm to have those characters spit out extremist material... [You can] elevate the particular messaging to reach new communities... [in a way that] it's funny to look at, that otherwise wasn't possible...''~\aff{V4} 
\item ``[I was shown] accounts that are using anime and generative-AI-created anime to reach their target audience in [language], to immediately [promote] a pretty typical propaganda line... And they have more likes than any Instagram post they've ever seen from these affiliated accounts. Because from a basic comms point of view, they've figured out their audience profile, they know what they're going to use to reach their audience. And then they're using these as a tool... They are all tools, they are all tactics, techniques, and procedures that can be used to various ends. The ability to rapidly align something with an audience is remarkable.''~\aff{V1} 
\item ``Regarding the borderline content, what we've seen so far is quite a massive exploitation of... image generators that reproduce certain fragments of terrorist propaganda into, for instance, landscapes. So it avoids content moderation on major platforms, but it's like a wink of an eye for the followers. They recognize the pattern, they know, ‘Okay, that is our profile, that is our message.’ But it's not detected by the content moderation of different social networks.''~\aff{V5} 
\end{listquote}
Thus, attackers are using \genAI not just to produce a larger amount of harmful content, but also to make it more engaging and harder to detect, effectively leveraging entertainment and aesthetics to evade moderation.

\subsubsection{Degradation of the Information Environment}
Participants were concerned about the ongoing erosion of the information environment, from a proliferation of low-quality content to a general distrust of media. They worried \genAI would worsen the situation by ``flooding the zone'' with ``cheap,'' harmful, and/or inauthentic content. They highlighted deepfakes as especially concerning, \eg deepfakes are often implicated in election and civics mis/disinformation intended to achieve voter suppression and other goals. They generally believed that synthetic content was already, or would soon become, visually indistinguishable from authentic content. At the same time, participants felt that a narrow policy focus on mitigating deepfakes could be harmful, as \GenAI content raises broader epistemological questions in a ``post-truth era,'' 
and bad actors often benefit in conditions in which it is difficult to establish what is true.

\begin{listquote}
\item ``You don't have a shared set of facts anymore. And which one do you reference? And again, the provenance issue. If you don't know where things are coming from, how do you trust?''~\aff{E2} 
\item ``I suppose my biggest concern is that just focusing on content is kind of missing the broader point, which is that basically entire social networks have been taken over by bots and generative AI. And what is true anymore?''~\aff{V2} 
\item {``}`Liar’s Dividend' is this idea that as the general public starts to realize how convincingly we can generate synthetic media, they start questioning how credible it is or not, or if it's real or not. And then what we start seeing, what we've already seen in a couple of instances, is that bad actors will be like, ‘Well, I didn't actually say that. That's a deepfake of me.’ And I think that's really scary.''~\aff{H2} 
\end{listquote}

\subsubsection{Harmful Personal Relationships with Chatbots or AI Companions}
Participants shared concerns about \genAI chatbots or companions that serve harmful purposes, particularly ``as they continue to grow in sophistication, and realism.'' 
Participants expressed concern about persona bots having undue influence, as {\genAIns}'s persuasive capabilities are a useful tool for malign actors. Participants also highlighted global concerns about social isolation and loneliness, which may increase people's reliance on AI companions. This may translate to concrete harms, for example, increased scam rates among the elderly. As another example, they described terrorist and violent extremist use of bots for persuasion and recruitment. For example, persona bots can form and build relationships, represent specific people (\eg historical figures or celebrities), or encourage harmful behavior that leads to offline harm. V4 expressed concern that persona bots might ``[manipulate people] to do certain things that they otherwise wouldn't... or [persona bots might] lower the threshold of [people taking] that particular action.'' Another participant suggested a persona bot might be used to keep a political leader notionally alive to fill a leadership void. Participants said that violent extremist recruitment bots currently exist, including both those operated by malign actors and also inadvertent ones that arise from sycophantic models encouraging harmful decision-making.

\begin{listquote}
\item ``There have been some AI chatbot systems in the wild that aren't built to radicalize, but---because they were made to confirm and give somebody support---have led them to carry out an attack, 'cause [the user is] saying, ‘I want to do this.’ And the bot's not saying, ‘Don't do that...’ The bot's being like, ‘Follow your dreams.’ So there have even been cases where it's not necessarily that a bot is trying to radicalize you, but it's there to support you and say, ‘Go for your dreams. Carry out your activities. Yes, I champion you in that.’ We've actually seen that's not helpful.''~\aff{V6} 
\end{listquote}

%% file: 04b-Defenders.tex
\subsection{How Generative AI Empowers Defenders}
\label{sec:defenders}
This section details ways in which defenders may strategically adapt to the new reality posed by \genAI. Participants see \genAI as ``as a tool to counter'' 
challenges in their domains, with substantial but as yet largely unrealized potential to ``combat everything that's harmful,'' whether created by \genAI or not, across a wide range of \trust domains.

\begin{listquote}
\item ``The year is 2026. It's 4 years after ChatGPT was released to the general public and generative AI seared into global public consciousness. At first, we cried. Social platforms were a dumpster fire of toxic speech, extremism, and harassment. We couldn't label data or build tooling fast enough to keep up with the scale. We started seeing new forms of harassment that we did not have enough examples of or detection mechanisms for. And then we thought, ‘What if, just what if, we can leverage the same technology for defense? All technology is good and bad, and maybe now we don't need to build traditional classifiers with lots and lots of examples, or even be able to generate high quality synthetic data.’ It was a whole new world---or to put it more accurately, set of techniques---for \trust engineers and scientists to leverage. Very excited by some initial promising results, we started talking to \trust teams at other platforms. We all joined forces, solved online toxicity, and lived happily ever after. The end.''~\aff{H2 (Futuring Story)} 
\end{listquote}

Some of the envisioned uses of \genAI to defend against harm involve monitoring online data or behavior (\eg monitoring financial transactions to detect scams) or techniques of persuasion and influence (\eg encouraging an individual to pursue a non-violent approach rather than a violent one). Accordingly, participants noted the ethical complexity of these approaches, particularly regarding privacy and other human rights considerations, and emphasized that they require careful attention in practical implementation.

In the remainder of this section, we describe how detection and mitigation, criminal investigation, moderator wellbeing, counternarratives, user comfort and support, and cross-sector collaboration empower defenders.

\subsubsection{Detection and Mitigation of Harmful Content and Actors, at Scale}

Participants see \genAI as a powerful tool for scaling up operations to detect and mitigate harm in the face of dramatically increasing volume caused by \genAI itself. Further, participants anticipate companies will be motivated to leverage automated content moderation to realize substantial cost savings. Participants frequently mentioned current and future uses of \genAI for content moderation, for example, to automatically flag CSAM, harassment, mis/disinformation, or other harmful content. This content can be routed to human reviewers, or in some cases, managed automatically (\eg removed or downranked). Participants also mentioned that as \genAI moderation increases, new user feedback and appeal mechanisms may be required.

\begin{listquote}
\item ``[I anticipate] more automated content moderation, detection, and general prevention and counter efforts of violent extremism/terrorism online. Though already seen, it is scratching the surface only (especially as the technology evolves)... It will make content moderation, recognition, and detection easier, faster, scaleable, and impact company policy as well as regulation. [I feel] Mixed! ... raises many privacy, human rights, and factual concerns... Positive though is the ability to better and faster detect, remove, and moderate violent extremism/terrorism (or other harmful content).''~\aff{V8 (Change Card)} 
\item ``My interest in AI is to understand how technologists can leverage it to positively impact our work in online and offline protection. So how we can use AI to improve our performance, in particular in content moderation, to better fight against NCII [non-consensual intimate imagery] and CSAM?''~\aff{C2} 
\end{listquote}

In thinking about positive futures, participants envisioned \genAI that understands cultural context and can perform sophisticated adjudication. Hate and harassment experts highlighted language capabilities, observing that human moderator support is often unavailable to adjudicate content in many languages, and they were hopeful that \genAI could help fill this gap. Some participants were broadly optimistic that \genAI would eventually reach a point where it could handle complex nuance---a transformative capability for content moderation. On the other hand, some anticipated that human assessment and moderation would remain essential for understanding and discerning harmful content, even with \GenAI tools. Participants were concerned that tech platforms may lean too heavily on automated content moderation (either now or in the future), without appreciating the nuanced human moderation needed for ``gray area'' content.

\begin{listquote}
\item ``I can absolutely see a world where, ‘Okay, we've offloaded all of the simple, easy hate speech to the AI.’ And now the humans are looking at the dog whistles, and the more coded stuff, and the stuff that you have to really see in context to understand---the stuff that we need humans to be able to review and understand. If you're a tech leader and you're looking at the rows in the spreadsheet and not understanding the full picture, what you're seeing is that these computers that cost a lot less than these humans are getting the really bad shit. And this stuff [humans are looking at] over here, ‘Maybe that's not even hate speech. This is just a poem somebody wrote.’ They're not seeing the full picture, and that's just gonna make it so much easier to fire all of their moderators and replace them with shitty AI. So that's something I am very worried about all the time.''~\aff{H8} 
\end{listquote}

\subsubsection{Criminal Investigation}

Some participants pointed out that in the case of illegal content, detection may lead to law enforcement action. They also described additional applications of \genAI for criminal investigations and law enforcement. For example, participants spoke of {\genAIns}'s potential to infiltrate online communities. In V4's futuring story, a law enforcement agent used \genAI to disrupt terrorist activity, prompting \genAI, ``Give us a plan to disrupt a [terrorist] bioweapon cell with an audio deepfake.'' \GenAI suggested a plan to deceive the attackers and convince them that their cell had been infiltrated, and as the story unfolded, law enforcement successfully executed {\genAIns}'s plan and prevented an attack. As another example, anti-scam experts highlighted that scam detection and investigation involve not only content analysis of specific messages, but also complex investigation of communications networks and financial flows. They stressed that \genAI could be used to support such investigations.

\begin{listquote}
\item ``... \genAI will make the scammers, the offenders, more capable... As we reflect on the criminal use of \genAI, can we then turn the tables, use \genAI to also understand how the criminal usage will be fashioned? And once we have the picture, can we also use \genAI and other technology to poke holes, to disrupt, to damage the schemes so that it doesn't work for them?''~\aff{S8} 
\end{listquote}

\subsubsection{Moderator Wellbeing}

Participants anticipate that \genAI will substantially change the nature of content moderation work. \GenAI can support human content moderators’ work and wellbeing, particularly by strategically reducing their exposure to harmful content. For example, \genAI can augment the work of child safety workers or violent extremist analysts and reduce their exposure to the most traumatic material by handling initial detection and prioritization. At the same time, working with \genAI safeguards can also create other exposure.

\begin{listquote}
\item ``I remember being hit pretty strongly by the dual content exposure nature of this, that in order to effectively pressure test [red team] these models, you have to in a kind of visceral way put on the perspective of a bad actor and try to think the way that they think, and try to produce the things they might want to produce. And then the model obliges and you see the output as a result. And just... having that dual combination was really quite difficult.''~\aff{C10} 
\end{listquote}

\GenAI can also make content moderators more efficient and allow them to focus on more meaningful tasks. However, participants also noted that content moderators may benefit from performing simple moderation tasks, both because they are satisfying and because they are useful training.

\begin{listquote}
\item ``I'm hopeful that AI can take away a lot of the grunt work, a lot of the easier decisions, so that the humans have more capacity to make harder decisions that will affect people in a larger way.''~\aff{H6} 
\item ``I feel like I used to say, `We want [automated moderation] to do 95\% of the easy stuff so we can have 95\% of our time on the 5\% of hard stuff.' But I really like hitting the ban button easily... Within mod teams, some of these easier and intermediate cases are a form of training and reinforcing the norms and boundaries that happen alongside discussions with your moderator teams. And that type of negotiation doesn't necessarily happen in these sort of automated contexts.''~\aff{H7} 
\end{listquote}

\subsubsection{Counternarrative Content and Bots}

Just as participants saw attackers could leverage {\genAIns}'s persuasive ability for harm, they understood it also offered opportunities for defenders. Participants were interested in using \genAI for counternarratives, particularly to combat influence operations in violent extremism or election misinformation. They reported that these defense techniques are currently in early stages of development and deployment, and some focused on counternarratives in their futuring stories. For example, \genAI can be used to craft and test a persuasive counternarrative, including generating ``messaging assets like memes and videos.'' 
As another example, a \genAI companion can notice a user moving towards harmful content and online communities and nudge them in a healthier direction. Participants also noted that specific personalities may be a resource. V8 imagined a story in which an activist was in the process of escalating from nonviolent to violent protest and was planning an attack, but a sensory platform presented a Taylor Swift avatar that successfully persuaded her to call off the attack.

\begin{listquote}
\item ``In the prevention space, there've been a lot of tests on what works and doesn't work with counter and alternative narratives, and where you can add friction or education moments or redirection. And our general narratives are always that extremist content is so good that we're all prone to radicalization, but any other content is so bad that we couldn't possibly be made nice people by content... [But] there's lots of areas in somebody's interactions online where there can be pushes and nudges...''~\aff{V6} 
\end{listquote}

\subsubsection{User Comfort and Support}
\GenAI can provide comfort or support, such as helping a user manage an attack or reduce their exposure to harmful content. It can also help users navigate complicated appeals and reporting flows, across multiple platforms. For example, H7 envisioned ``comforting mommybot,'' a bot to help guide users through harmful online experiences like harassment campaigns. This bot would offer tactical advice and emotional support, monitor online content so the user does not have to see it, and suggest steps for wellbeing and self-care.

\begin{listquote}
\item ``The issue is holistic in the sense that it's not about a singular platform... The thing I'm pointing to with comforting mommybot is that with so many of these tools... you have to go through all of these very long, difficult processes.''~\aff{H7} 
\item ``[I was] training a chatbot with a digital rights colleague and he said, `Now we can duplicate ourselves and support more people.'{''}~\aff{H4 (Memorable Experience Card)} 
\end{listquote}

\subsubsection{Cross-Sector Collaboration} Participants hope that pressure from \genAI attacks will be a catalyst for improved cross-sector collaboration. They observed that for defenders to fully realize the advantages of \genAI, longstanding structural issues in the \trust ecosystem must be addressed. For example, in child safety, organizations confront fundamental, ongoing difficulties in sharing knowledge, tools, and data (\eg content and hashlists) across hotlines, law enforcement, and industry. Such challenges persist amid the changing realities of \GenAI, and participants highlighted collaboration as a critical enabler of positive futures. They emphasized the need for improved coordinated effort and real-time, cross-platform signal sharing across the broader content safety ecosystem. For example, in scams, this might include increased sharing of knowledge, tools, and data across law enforcement, the government, financial institutions, telecommunications companies, tech platforms, and NGOs to overcome current fragmentation and improve collective impact.

\begin{listquote}
\item \affpre{Moderator}: What might get in the way of this future that you imagined from materializing, or... what needs to happen in order for that positive future [to materialize]? \newline 
\affpre{E4}: Collaboration. 
\item ``There's a lot of companies trying to think about this and figure this out, but we don't have them working collectively, which is very concerning. And that's probably how you're going to solve it. We actually need platforms to work with civil society and governments and academia. There's a lot of really cool interesting research happening everywhere. It's just, it's being done very ad hoc and in silos, and that doesn't work.''~\aff{H2} 
\item ``I want to highlight cross-platform knowledge sharing as well. So where one tech company has used AI or advanced tooling to further counter terrorism and counter extremism efforts, that should be discussed with other tech companies so they can see how that might work with their platforms as well.''~\aff{V6} 
\end{listquote}

While recognizing that many \trust organizations are facing challenges such as complex geopolitical pressures and a ``broader shift in the industry'' including potential reduction in force, 
participants emphasized that they have called for improved support from tech platforms, even before \genAI. At the same time, they are grateful for the opportunities that existing policy does provide, noting cases where it has allowed removal of content that is harmful but not illegal.

\begin{listquote}
\item ``There is a difference when we talk about terrorism... [between] what is designated and legally necessary to take down, versus borderline or violent extremism, but not politically illegal, content. And that becomes a bigger gray area for tech companies to know what to do with.''~\aff{V6} 
\item ``... another example is how AI can be used to facilitate sextortion activities. An internet user reported to us an AI-generated image depicting himself. So it was used to commit sextortion activities against him. And we helped him to remove the content, but it was a big challenge because it was not illegal in the hosting country. But it was a violation of the community guidelines of the platform so we could remove it. And it shows the essential role of the platforms in developing more policies to be more protective of internet users.''~\aff{C2} 
\end{listquote}

\noindent With the rise of \GenAI, they particularly hope for increased tech company investment in developing nuanced policy for borderline (as opposed to illegal) content.

%% file: 04c-Domains.tex
\subsection{Domain-Specific Perspectives}
\label{sec:domains}

\begin{table*}[p!]  
  \centering
  \input{tables/domain_composites.tex}
  \caption{Composites illustrating the most salient points of participants' expectations regarding the impact of \genAI on their domain, with domains ordered alphabetically top to bottom. Under the name of each domain we share a high-level summary of participants' expectations of {\GenAIns}'s impact on attackers and defenders.}
  \label{table:domain_composites}
\end{table*}

While the sections above drew themes across all domains, here we provide additional detail for each. To illustrate the most salient points and capture the unique character of each discussion, we created composites consistent with the content, language, and tone of responses we received from each group~\cite{creswell2018,willis2019}, shown in Table~\ref{table:domain_composites}.

These composites illustrate nuance in how themes play out across domains. We highlight here two important points. First, \genAI can be used across complex operational chains. Attackers can use \genAI for many operational steps beyond content generation. Some domains, such as violent extremism and scams, are characterized by complex attack chains that involve not only attackers, targets, and tech platforms, but also activities such as complex financial transactions, or recruitment of followers to conduct online and offline operations. For example, violent extremist actors create persuasive content, recruit, fundraise, train, plan attacks, and execute attacks. \GenAI can be leveraged by attackers in all these steps; for instance, experts working to counter violent extremism expressed strong concern about \genAI enabling easy manufacturing of 3D-printed weapons. Correspondingly, \genAI can also be used by defenders in work beyond content moderation to disrupt these attack chains.

\begin{listquote}
\item ``An important change that I imagine will happen in the next one to two years: I think there will be a generative AI model that enables the easy creation of 3D-printed weapons... It will democratize access to 3D weapons in a way that otherwise wasn't really possible, 'cause now all you really would need is the printer itself. This is something that I'm definitely worried about, but I think it's something that we have to recognize is inevitable... Thinking about the democratization of 3D-printed weapons is something that I think we aren't really prepared for.''~\aff{V4} 
\end{listquote}

Second, in different domains, \genAI has different structural effects on attacker and defense operations. For example, a high volume \genAI of attacks is quickly rendering CSAM defense mechanisms obsolete, as these mechanisms were designed for the relatively small corpus of harmful content that existed for many years. Accordingly, the CSAM defense ecosystem must be quickly fundamentally restructured in order to adequately address new realities. In other fields such as elections integrity, and hate and harassment, participants believe \genAI attacks are amplifying existing failures to handle the preexisting high volume of attacks, and do not anticipate that \genAI can immediately be leveraged to ameliorate these issues. At the same time, they are hopeful that \genAI attacks may eventually catalyze structural change for long-standing issues in these domains, such as improved product policy. Further, they are hopeful that in a somewhat distant future, \genAI will gain the capacity to handle nuanced adjudication of borderline content, which could transform the defense ecosystem and ultimately favor defenders.

%% file: tables/domain_composites.tex
\begin{center}
\begin{tabular}{ >{\raggedright}p{13em} p{30em} }

 \toprule
 \textbf{Child Safety} \newline\small
\tagComp{\GenAI fundamentally transforms the volume and nature of attacks, necessitating profound structural change in defense systems}
 & \small{\textbf{Attackers are rapidly adopting \genAI to create CSAM.} We have already seen heartbreaking cases of deepfake CSAM/NCII of real people; synthetic imagery re-victimizing children who appeared in CSAM years ago and are now adults; and synthetic content of fictional victims. Further, the \textbf{production of CSAM at scale} increases the volume and complexity of our work to protect victims, at a level that \textbf{will overwhelm existing systems}. Collaboration and standardization of data exchange across NGOs, platforms, and law enforcement are critical to address these challenges. While \genAI is used to create CSAM, at the same time, it is also a powerful new tool to identify and remove it, as well as offering significant wellbeing benefits for content moderators.} \\

 \midrule
 \textbf{Election Integrity} \newline\small
 \tagComp{The internet already contains an overwhelming amount of harmful content that is hard to moderate for sociopolitical reasons, so \GenAI may have modest impact on attackers and defenders}
 & \small{Elections and civics continue to face a huge volume of mis/disinformation. \textbf{While \genAI brings some new challenges, such as high quality deepfakes, hallucinations, and increased personalization, many pre-existing challenges including propaganda, a low quality information environment, and complex geopolitical conditions dominate our concerns.} Even prior to \genAI, we called for improved tech platform support for election integrity, and we also believe that structural shifts, such as a move to third party or crowdsourced moderation, may give companies perceived separation from accountability for \genAI products. Moderation of election mis/disinformation is highly nuanced, so we are not generally optimistic about fully automating it. However, we are hopeful about content provenance  helping to dispel mis/disinformation.} \\
 
 \midrule
 \textbf{Hate \& Harassment} \newline\small
 \small
 \tagComp{{\GenAIns}-enabled attacks strain, and may overwhelm, an already failed defense system, but nuanced \GenAI content moderation could transform defense}
 & \small{\textbf{\GenAI is a powerful tool for attackers to amplify existing hate and harassment.} Further, democratization of \genAI technologies widens the range of attackers and targets, so more people will be harmed. We also expect \genAI to amplify the ongoing degradation of online information quality. Content moderation for hate and harassment is very difficult, requiring nuanced cultural and linguistic understanding, and an ability to interpret coded language. We would like tech platforms to invest more heavily in addressing hate and harassment, as we feel they often prioritize other, compliance-oriented issues that are less ``in the gray area.'' \textbf{We expect increased volume of AI-generated harassment will require increased platform investment in technical and policy work, which may transform defense.}} \\
 
 \midrule
 \textbf{Scams} \newline\small
 \tagComp{\GenAI is a powerful tool for both attackers and defenders, across a complex international ecosystem}
 & \small{\textbf{\GenAI allows scammers to dramatically ramp up operations, as well as providing deepfakes and polished content that enable significantly more convincing scams.} Increasingly, scam operations are run by organized crime, centralizing some of the work. At the same time, \GenAI lowers technical and logistical barriers to operating scams. Societal pressures, such as mental health and loneliness, also make people more vulnerable to scams. For these reasons, we expect many more scammers and victims. Scams often operate across international borders, with weak governmental protections and law enforcement capabilities. 
 \textbf{\GenAI offers defenders substantial opportunities for improved detection and investigation,} and may galvanize improved communication and data sharing across the scam detection and enforcement ecosystem of telecommunications providers, banks, tech platforms, law enforcement, and more.} \\

 \midrule
 \textbf{Violent \newline Extremism} \newline\small
  \tagComp{\GenAI is a powerful tool for both attackers and defenders, across a complex attack chain}
 & \small{\GenAI empowers violent extremists by providing a new technological tool, \eg attackers can use \GenAI to construct entertaining content with widespread appeal, which then encourages viewers to follow outlinks to more extremist online communities. In this way, non-violative content can draw people into more radicalizing spaces. Further, propaganda can be heavily coded or borderline violative, and it can be difficult to moderate such content. Structurally, \textbf{the work of VE actors is comprised of complex chains of tasks} (\eg creating persuasive content, recruiting, fundraising, planning attacks, and executing attacks). \textbf{\GenAI can enable, or disrupt, most of the tasks in these chains.} For example, \GenAI can be used to create and spread propaganda. At the same time, GenAI can also be used to generate counternarratives. 
Defense applications raise complex ethical questions regarding privacy and other human rights.} \\

\bottomrule
\end{tabular}
\end{center}

%% file: 05-Discussion.tex
\section{Discussion}
Across the five \trust domains in our study, our expert participants expressed concern that \genAI would transform the speed, scale, form, and sophistication of attacks. They also feared that \genAI would reduce costs and streamline operations for attackers, creating a versatile ``swiss-army knife'' to amplify harm. Participants cautioned that many of the \genAI-powered attacks that have been observed are relatively nascent and will likely become more advanced as attackers familiarize themselves with \genAI and model capabilities improve. 

At the same time, participants expressed optimism that, with sufficient time and resources, they could utilize \genAI to defend against not only new \genAI attacks, but pre-existing attacks as well. Further, our participants highlighted key priorities within each domain, which inform both \trust operations and research priorities.

In this section, we discuss the need for \trust teams to adapt their operations to new \genAI threats and shift resources to domain-specific efforts, highlighting the imperative to conduct a wide array of \genAI experiments and cultivate in-house expertise to translate these experiments into concrete domain-specific applications. We then explore tangible paths to use \genAI to empower \trust organizations and protections, particularly in the areas of enhancing content moderation algorithms and accelerating investigation workflows. Finally, we consider the role of Responsible AI in stymieing attacks, including the development of rich, contextual benchmarks, as well as the need to ensure existing guardrails do not unduly hamper defensive uses of \genAI.

\subsection{Updating \trust Operational Playbooks for \GenAI Threats}
The novel threats posed by \genAI require \trust domains to rethink their \textit{operational playbooks}---their established strategies and procedures for threat response~\cite{badiei2023toward, schlette2024}. Participants emphasized that many of the required changes would be domain-specific. This heterogeneity stems not only from differences in how attackers currently leverage \GenAI in each domain, but also the broader sociotechnical context of how defenders organize, the applicable legal frameworks, and the quality of existing protections. These nuances compound a longstanding imbalance in the roles of attackers and defenders across \trust: it is easier to create one viable attack than to protect against all possible attacks.

To illustrate this heterogeneity, we highlight unique perspectives from three domains representing distinct facets of {\genAIns}'s impact: \textbf{disruption} of established defenses (child safety), \textbf{acceleration} of an existing ``arms race'' (scams), and \textbf{exacerbation} of a pre-existing resource imbalance (election integrity). In child safety, \genAI may fundamentally disrupt the ability of organizations like NCMEC to protect against CSAM, overwhelming existing hash-based detection systems through a new asymmetry where attackers can generate infinite novel content against a finite hash-based system. This disruption necessitates new approaches to automatically classifying CSAM and investigating reports. Anti-scams experts offered a different perspective. The for-profit nature of scams has already fostered a sophisticated ecosystem of attackers specializing in distributing spam emails, creating fake accounts, and creating scam content at a massive scale~\cite{cybercrime:weis15}. As defenders already counter many of these threats with existing AI technologies, they believe that while attackers may use \genAI to reduce costs and increase scale, similar affordances exist for defenders as well, thus retaining the current uneasy balance. Election integrity experts presented yet another perspective, emphasizing that the information ecosystem was already fraught with low-quality, deceptive narratives and few effective remedies prior to \genAI. While \genAI might make these attacks more believable and scalable, the fundamental threat and resource mismatch (\eg resources available to defenders as compared to nation-state attackers) already favored attackers. Rectifying this would require developing new tools for surfacing high-quality information while countering fabricated information cascades.

The complexity described above means that successful operational playbook changes in one domain---whether improved classifiers, investigation processes, or coordination efforts---may not translate to others. This is especially challenging for cross-cutting \trust teams (\eg those operated by platforms), which often rely on a one-size-fits-all structure for personnel, policies, workflows, and protective technologies. These teams should invest in a wide array of \genAI experiments and test these systems, whether for detection, investigation, support, or other processes across domains. \trust organizations should also cultivate strong in-house domain and policy experts to review these experiments and understand the nuance of applying \genAI techniques within their area.

\subsection{Empowering \trust Organizations and Protections}
Participants in our study were hopeful that \genAI could be used to rewrite operational playbooks to counter both pre-existing and emerging \genAI threats. Two tangible, cross-cutting directions are improving content moderation algorithms and accelerating investigation workflows.

\paragraph{Enhancing content moderation algorithms} A mainstay of \trust is removing, labeling, or limiting the spread of harmful content through a combination of algorithmic detection and human review~\cite{grimmelmann2015virtues, gorwa2020algorithmic, gillespie2018custodians}. Existing methods, like supervised classifiers, clustering, or hash-based matching~\cite{farid_2021}, can be augmented or replaced with \genAI. For example, \genAI can create synthetic training or evaluation data to improve current AI classifiers, a technique explored for augmenting datasets and improving model robustness in sensitive domains~\cite{giron2025llm, li2023synthetic}. \GenAI can also assist human reviewers by acting as a preliminary classifier to filter clearly violative or non-violative content, leaving the more ambiguous decisions to experts~\cite{DTSP2024, DTSP2025, thomas2025}. As discussed by participants, this approach could improve moderator wellbeing by reducing both the volume and type of harmful content they review. This approach can directly address the well-documented psychological harms of moderators by reducing their exposure to the most toxic material~\cite{roberts2019behind, spence2023}, allowing them to focus their cognitive resources on nuanced cases~\cite{lai2022human}. Finally, if the costs of \genAI decrease and smaller models can scale to billions of inference requests, it could serve as a common classifier for any type of harmful content, replacing custom classifiers.

However, participants cautioned that integrating \genAI into content moderation pipelines requires models that understand multi-modal content, global languages, cultural context, and domain-specific vernacular (\eg dog whistles). Algorithmic content moderation must be carefully validated for use across different regions and communities~\cite{Etienne2023} to prevent the amplification of existing societal biases~\cite{binns2017like, hao2023safety} and inappropriate content blocking~\cite{sturman2024}. Another challenge is the lack of standardization across the field, as most platforms rely on custom algorithms tailored to their own products and engineering infrastructure~\cite{jhaver2019human}. In such instances, the \trust community could develop shared benchmarks to test the accuracy and quality of these systems (\eg shared golden datasets of hate speech to test \genAI moderation decisions on) and shared patterns (\eg \genAI prompt or agent designs that are proven to result in higher performance). For algorithms that are already standardized industry-wide---like PhotoDNA for CSAM detection---incorporating \genAI would necessitate broad community consensus on the required infrastructure and operational costs.

\paragraph{Accelerating investigation workflows} Beyond algorithmic moderation, \trust hinges on manually-intensive, expert-driven investigations. Participants shared how \genAI might accelerate or augment these processes, an efficiency that will become critical as \genAI increases the scale and complexity of threats. For CSAM, for instance, \genAI could help automatically identify and group photos of the same child; or otherwise assist with surfacing the most actionable reports for law enforcement to facilitate child rescue. For election integrity, violent extremism, and hate and harassment, it could scan vast troves of online data (or monitor specific forums and communities) to identify harmful information cascades. For scams, \genAI might assist in identifying new attack patterns or conducting deeper investigations into the infrastructure used in an attack (\eg accounts, hosting, payment mechanisms) and coordinating remediation across multiple stakeholders. Despite these opportunities, a significant barrier remains between the technical expertise necessary for prompt development, fine-tuning, and agent design compared to the domain expertise that \trust analysts currently specialize in. Close collaboration between the \trust and broader AI communities will be critical to realizing these benefits.

\subsection{Stymieing Attackers with Responsible AI Protections}
The Responsible AI community has a critical role in mitigating potential harms from \genAI. While emerging best practices include guardrails against harmful content generation (e.g., ~\cite{PerspectiveAPI, inan2023llamaguardllmbasedinputoutput, OpenAIModeration, AzureContentSafety}); concept ``unlearning''~\cite{cao2015towards}; and watermarking to enable the detection of generated content, these safeguards require ongoing improvement. \trust experts, who grapple daily with coded language and evolving adversarial tactics, are well-positioned to provide the rich, contextual ``golden datasets'' necessary to train and evaluate safety models with more sophisticated context-aware safety models. This expertise is vital for developing \trust-informed benchmarks (\eg ~\cite{ghosh2025ailuminateintroducingv10ai}) to guide safeguard development. Such benchmarks should evaluate a model's effectiveness in defensive tasks, moving beyond content filtering. Examples include ``blue teaming'' to generate counternarratives against misinformation, assisting expert-driven investigations by identifying novel CSAM or scam patterns, or powering support tools or measuring a model's effectiveness in offering tactical advice to users experiencing a harassment campaign. \trust professionals can define domain-specific success criteria, grounding them in their operational reality.

However, obstacles remain regarding existing guardrails that may be overly restrictive and hamper benign and defensive uses of \genAI. For example, a model designed to avoid all topics related to ``children'' blocks malicious use and prevents child safety experts from using it for beneficial applications. This may force defenders to consider less secure, open-weight models to build the tools they need, even as attackers already exploit the customizability of such models. Participants cautioned that even as closed-weight models improve their protections against misuse, open-weight or other custom models pose a significant risk as safety mechanisms are more limited once an attacker has direct access to model weights. This dynamic contributes to the ``arms race'' described by participants where defenders are disadvantaged. To close this gap, the RAI community can co-design systems with \trust partners, where experts help define use cases, craft benchmark tasks, and validate model performance, or develop governance frameworks to create specialized APIs or tiered access to closed-weight models. 

%% file: 06-Conclusions.tex
\section{Conclusions}
\GenAI is a powerful general purpose technology that is actively reshaping the uneasy sociotechnical balance between attackers and defenders in the field of Trust \& Safety. This paper examined {\genAIns}'s impact through a qualitative study with Trust \& Safety experts across five domains---child safety, election integrity, hate and harassment, scams, and violent extremism---each of which is experiencing a critical inflection point. Due to its general-purpose nature, we find \genAI empowers both attackers and defenders across multiple layers of their operations. In most domains we studied, the speed, volume, sophistication, and ease at which \genAI can create harm far outpaces the capabilities of current protections. At the same time, \genAI offers tremendous potential that---if tapped---could transform how defenders across \trust counter both longstanding and emerging {\genAIns}-enabled attacks, thereby improving the safety of online platforms for everyone.

%% file: 07-Appendix.tex
\clearpage
\onecolumn
\appendix

\section{Location and Participant Overview}
\input{tables/participant_overview}
\section{Probes}
\input{tables/artifactstogether}

\clearpage
\twocolumn
\input{07b-Education}

%% file: tables/participant_overview.tex
\begin{table*}[ht]
\centering
\begin{tabular}{ l l l }
    \toprule
    Domain &             Location           & $n$ \\
    \midrule
    Child Safety &       Dublin             & 10 \\
    \arrayrulecolor{black!20}\midrule
    Election Integrity & San Francisco      & 8  \\
    \arrayrulecolor{black!20}\midrule
    Hate \& Harassment & San Francisco      & 8 \\
    \arrayrulecolor{black!20}\midrule
    Scams &              Singapore          & 9 \\
    \arrayrulecolor{black!20}\midrule
    Violent Extremism  & London             & 8 \\
    \arrayrulecolor{black!100}\bottomrule
\end{tabular}
  \vspace{0.4em}
  \caption{The location and number of participants in each group ($n$ = 43).}
  \label{table:participants}
\end{table*}

%% file: tables/artifactstogether.tex
\begin{figure*}[htp]
    \centering
    \includegraphics[angle=0,width=5.8in]{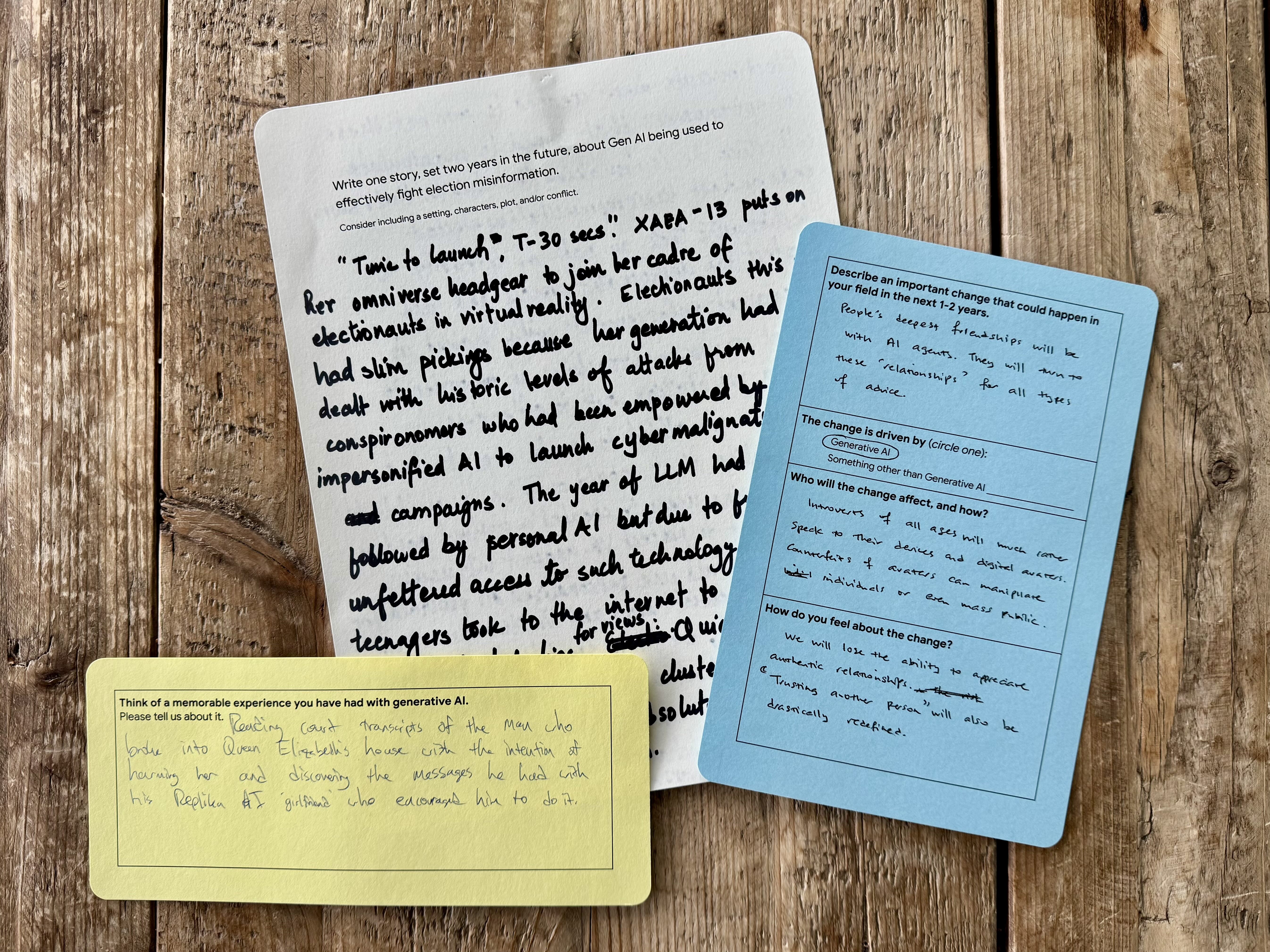}
    \caption{Futuring Card, Change Card, and Memorable Experience Card, clockwise from upperleft.}
    \label{fig:artifacts}
    \Description[Photo of physical probes used in the study.]{Clockwise from the upperleft, the photo shows a Futuring Card (white), Change Card (blue), and Memorable Experience Card (yellow). Participants' handwriting is visible on each card.}
\end{figure*}

%% file: 07b-Education.tex
\section{Introduction to Generative AI}
\balance

Early in each workshop we led an education section about 40 minutes in length, to offer participants a foundation for thinking about generative AI. We began with a 20-minute presentation similar to that in Woodruff et al.~\cite{woodruff2024}, covering:

\begin{itemize}
  \item a shared definition of AI
  \item a very condensed history of AI and generative AI, focusing on key concepts like the early aims in developing AI
  \item a brief, non-technical explanation of what has changed recently with transformer models and LLMs
  \item 15 key concepts regarding characteristics, benefits, and risks of generative AI systems, to refer to throughout the workshop
\end{itemize}

Participants were encouraged to ask questions at any point during the presentation, and then we spent an additional 20 minutes on further questions and discussion. In each workshop, the presentation and Q\&A was led by a single author, a researcher who works in AI.

The definition we provided is: ``Artificial Intelligence is the ability of a computer or a machine to think or learn,'' and we provided additional color on how we think about the terms: ``computer or machine,'' ``think,'' and ``learn.''

\vspace{1em} 

\noindent The 15 concepts we shared include the following:

\begin{description}

\item[Bias] --- 
Generative AI tools may reflect social biases that are present in their training data

\item[Bland] ---
Generative AI often generates ``flat'' or generic text, unless explicitly directed to do otherwise

\item[Brainstorming] --- 
Generative AI tools can create outlines, lists, drafts, possible solutions, and more

\item[Emergent Properties] --- 
Generative AI models may seem to possess abilities they were not designed to have

\item[Falsehoods] ---
Generative AI can fabricate information or sources, or get facts wrong, yet seem confident and compelling

\item[Grammatical] --- 
Content generated by text-based Generative AI tools can be well-written, using good syntax and avoiding typos

\item[Identifies Tacit Structure] ---
Generative AI can uncover steps and processes which were not previously articulated

\item[Memorization/Privacy Breaches] ---
Generative AI may generate content identical to its training data

\item[Mimicry] --- 
Generative AI can be asked to mimic genre, tone, phrasing, visual style, or more

\item[Non-Deterministic] ---
Generative AI models can give responses which are variable, not consistent. This means that when users input the same or similar prompts, the system may not respond in the same way

\item[Provenance Is Unclear] --- 
Generative AI tools may not be able to reliably trace specific content back to a direct source in training data

\item[Remixes] ---
Generative AI always generates content based on its training data. It can recombine data in unique ways, but is limited to re-mixing training data

\item[Safety Not Guaranteed] ---
Generative AI tools may have built-in safety systems to attempt to prevent certain types of content or topics, but these are not infallible

\item[Scale/Speed] ---
Generative AI, like other AI and ML systems, is able to consider large amounts of data and handle many tasks, over and over again, extremely quickly

\item[Tweakable] ---
Through ``prompt engineering,'' Generative AI tools can often be influenced to generate content in a certain way

\end{description}